\documentclass[aps,prb,twocolumn,showpacs,superscriptaddress]{revtex4}

\usepackage{tabularx,graphicx}
\usepackage{amsmath, amsthm, amssymb} 
\usepackage{bbm}
\usepackage{dcolumn}
\begin{document}
\title{Magnetic structure of an imbalanced Fermi gas in an optical lattice}
\author{B. Wunsch}\email{bwunsch@physics.harvard.edu}
\author{L. Fritz}
\author{N.~T. Zinner}
\affiliation{Department of Physics, Harvard University, Cambridge, Massachusetts 02138, USA}
\author{E. Manousakis} 
\affiliation{Department of Physics and MARTECH, Florida State University, Tallahassee, FL32306-4350, USA}
\author{E. Demler}
\affiliation{Department of Physics, Harvard University, Cambridge, Massachusetts 02138, USA}
\date{\today}
\begin{abstract}
  We analyze the repulsive fermionic Hubbard model on square and cubic lattices with spin imbalance and in the presence of a parabolic confinement. We analyze the
  magnetic structure as a function of the repulsive
  interaction strength and polarization.  
 In the first part of the paper we perform unrestricted Hartree-Fock calculations for the 2D case and find that above a critical interaction strength $U_c$ the system turns ferromagnetic at the edge of the trap, in agreement with the ferromagnetic Stoner instability of a homogeneous system away from half-filling. For $U<U_c$ we find a canted antiferromagnetic structure in the Mott region in the center and a partially polarized compressible edge. The antiferromagnetic order in the Mott plateau is perpendicular to the direction of the imbalance. In this regime the same qualitative behavior is expected for 2D and 3D systems.
  In the second part of the paper we give a general discussion of magnetic structures above $U_c$. We argue that spin conservation leads to nontrivial textures, both in the ferromagnetic polarization at the edge and for the Neel order in the Mott plateau. We discuss differences in magnetic structures for 2D and 3D cases.
     \end{abstract}
\maketitle
\section{Introduction}
  Cold atoms  constitute a promising route to simulate {\it model} Hamiltonians of strongly correlated many-body physics with accurate control of system parameters\cite{bloch2008,Jacksch1998}. After major experimental breakthroughs with ultracold bosonic atoms like the Bose-Einstein condensation (BEC) of alkali gases \cite{wieman1995,ketterle1995} or the observation of the superfluid-Mott insulator transition in a bosonic Hubbard model \cite{greiner2002}, the field of ultracold atoms is currently addressing problems of strongly correlated fermionic systems\cite{LeHur2009,Giorgini08,Lewenstein07,Ketterle08}. Arguably the most prominent goal is the understanding of the phase-diagram of the fermionic Hubbard model, which is believed to be of major importance for high-temperature superconductivity \cite{hubbard, hubbardrev,Bruun,LeHur2009,hofstetter,Pollet,Sarma}.   
A two-component Fermi-gas in an optical lattice is well-described by the single-band Hubbard model, whenever the energy gap to higher bands is much larger than onsite interaction, temperature, and chemical potential\cite{bloch2008,Jacksch1998,hofstetter}. 
Only recently the fermionic Mott transition has been realized experimentally\cite{jordans2008,Mott2}.
The major challenge for studying magnetism of the fermionic Hubbard model is to reach temperatures below the N\'eel temperature\cite{koetsier2008,Neel2}.
In addition to the preparation of the antiferromagnetic state, characterization tools have to be developed to allow a clear identification of the magnetic structure.
Possible experimental techniques include Bragg spectroscopy\cite{Bragg1,Brag3}, local measurements of the magnetization\cite{vengalattore2007,Trotzky08},  noise correlations\cite{altman2004,Noise}, or the recently realized quantum gas microscope\cite{Greiner}.

The experimental control of spin imbalance in Fermi gases offered a unique way to study 
pairing phenomena beyond the standard BCS picture for attractive interactions\cite{zwierlein2006,partridge2006}. 
Motivated by these results, we address in this work the effect of spin imbalance on the repulsive fermionic Hubbard model\cite{Koetsier09,snoek2008}. 
While we study strong optical lattices, where a single-band Hubbard model is realized, the magnetic structure of weak to intermediate lattice strength including multiple bands has also been  discussed\cite{Mathy}.
We find rich physics arising from the interplay between antiferromagnetic and Stoner ferromagnetic instabilities and spin imbalance. 

The magnetic order of the two-dimensional repulsive Hubbard model has been extensively studied in the past, see e.g. Ref.\onlinecite{hubbardrev}.
Cold atoms in optical lattices differ in several ways from typical condensed matter 
systems. First, there is a superposed external confinement potential, which divides the system in an incompressible Mott state in the center of the trap and a compressible region at the edge. Second, the total spin is conserved, which means that we need to minimize the energy of the system given a global magnetization rather than a finite Zeeman field. One interesting problem concerns the spatial distribution of the imbalance between Mott plateau and edge, and it turns out that the solution strongly depends on the interaction strength.
The constraint of spin conservation affects the ferromagnetic instability at the edge by enforcing nontrivial spin textures\cite{Berdnikov09,LeBlanc09} which also affects the Neel order in the Mott plateau in the center, as we will  discuss in section~\ref{Topology}.  

In this work we study the repulsive fermionic Hubbard model including a parabolic confinement potential. In the first part of this work we perform unrestricted Hartree-Fock calculations for the 2D case.
Relevant physics for this system can be identified based on the mean-field phase diagram for the repulsive 2D homogeneous Hubbard model\cite{Hirsch85}. 
Up to a critical interaction strength $U_c$ it predicts antiferromagnetic order close to half -filling and paramagnetic order elsewhere.
In the spirit of a local density approximation one might then expect that cold fermionic atoms in an optical lattice have antiferromagnetic correlations in spatial regions with one atom per site and are paramagnetic elsewhere. In order to account for a finite imbalance, the system has to change its magnetic structure. Using an unrestricted Hartree-Fock approach for the 2D system we find a canted antiferromagnet in the Mott plateau in the trap center and a partially polarized edge. We note that canted antiferromagnetic order close to half-filling has been reported previously in Ref.~\onlinecite{gottwald2009}.
With spin polarization along the $z$-direction, the canted antiferromagnet accommodates the imbalance forming a constant $z$-component of the local magnetization, and simultaneously it benefits from the superexchange interaction by building up an alternating magnetic order perpendicular to the $z$-direction. Fixing the global imbalance and increasing the interaction strength results in more imbalance flowing to the edge. 

Above a critical interaction strength $U_c$  the unrestricted Hartree-Fock calculation predicts that the system turns ferromagnetic at the edge of the trap, in agreement with the ferromagnetic Stoner instability of a homogeneous system away from half-filling. Furthermore, the orientation of the antiferromagnetic order in the Mott plateau is perpendicular to the direction of the ferromagnet in the edge. Spin conservation has again a strong impact on the magnetic structure of the system, since a uniformly polarized ferromagnetic edge together with an antiferromagnetic Mott plateau are generally not allowed. We will discuss spin textures in 2D and 3D lattices for $U>U_c$, which fulfill spin conservation and which show the  two prominent features predicted by the mean-field calculation, namely a) magnetic instabilities towards ferromagnetism in the compressible edge and antiferromagnetism in the Mott plateau and b) at the interface between Mott plateau and compressible edge, the orientation of the antiferromagnet and the ferromagnet are perpendicular to each other.

We are aware that the chosen mean-field approach generally overestimates symmetry breaking and therefore the critical on-site interaction strength, $U_c$, corresponding to the appearance of an intrinsic ferromagnetic edge, will presumably be higher than the one predicted here. However, intrinsic ferromagnetism away from half-filling is expected for sufficiently large interaction strength\cite{Park08,Berdnikov09}
and in fact  experimental indications for itinerant ferromagnetism in a Fermi gas of ultracold atoms have been reported recently in Ref.\onlinecite{Jo09}.
Given the tunability of the ratio between  onsite interaction and nearest neighbor hopping, $U/t$, the interaction strength required for the presented phase separation should be accessible in experiment ($U/t=150$ have been reported in Ref.\onlinecite{jordans2008}).

The paper is organized as follows.  In Sec.~\ref{model} we introduce the model and in Sec.~\ref{HFsection}  we calculate the  magnetic structure for $U<U_c$ within an unrestricted Hartree-Fock approach. The topology of the intrinsically ferromagnetic edge arising for $U>U_c$ is addressed in Sec~\ref{Topology} and in the Appendix. Finally in Sec.~\ref{conclusion} we summarize our findings and comment on the experimental significance of our results.

\section{Model}\label{model}
We consider the fermionic single band Hubbard model on a 2D and 3D cubic lattice with an external parabolic confining potential. The Hamiltonian is
\begin{eqnarray}\label{eq:Ham}
H=-t \sum_{\langle i,j\rangle, \sigma} c_{i\sigma}^\dag c_{j\sigma} +U  \sum_i n_{i\uparrow} n_{i\downarrow}+\alpha\sum_i r_i^2 n_i,
\end{eqnarray}
where $\sigma\in\{\uparrow,\downarrow \}$ labels the two
fermionic components, which are the eigenstates of the $z$-component of a spin
algebra. These two components can either be the hyperfine state of the trapped fermions or even correspond to different atomic species.  $c_{i\sigma}$ denotes the annihilation operator for a
particle with spin $\sigma$ at site $i$, $n_{i\sigma}=c_{i\sigma}^\dag
c_{i\sigma}$ and $n_{i}=\sum_\sigma n_{i\sigma}$ are the spin resolved
and total occupation of site $i$.  $U$ is the on-site interaction and $t$ is the nearest neighbor hopping. Finally $r_i$ denotes the distance
of site $i$ from the trap center measured in units of the lattice
spacing $a$ and $\alpha=m\omega^2 a^2 /2$ characterizes the strength of the
external confinement. The associated energy scale is the confinement strength at the edge of the atom cloud with one atom per site, denoted by $V_t$ . In 2D $V_t=N \alpha/\pi$, where $N$ is the particle number.

\section{Unrestricted Hartree-Fock approach in 2D}\label{HFsection}
We now apply a Hartree-Fock mean-field decoupling in the spin and the density channel. Since the trap breaks translational invariance, the mean-field parameters will be site-dependent.
Allowing for arbitrary spin and density at each site we obtain the following mean-field Hamiltonian\cite{Verges91}
 \begin{eqnarray}
H&=&H_0+H_{int}\label{eq:MF2}\\
H_0&=&-t \sum_{\langle i,j\rangle, \sigma} c_{i\sigma}^\dag c_{j\sigma}+\alpha\sum_i r_i^2 n_i \,\notag\\
H_{int}&=&U  \sum_i \left(\frac{1}{2}  n_{i} \langle n_{i} \rangle-2 \vec{S}_i \cdot\vec{M}_i \right)\notag\,,
\end{eqnarray}
where $\vec{S}_{i} =(\sum_{\alpha,\beta}c_{i \alpha}^\dag \vec{\sigma}_{\alpha,\beta} c_{i \beta}) /2$ denotes the spin operator at site $i$ ($\vec{\sigma}$ is the vector of Pauli matrices) and $\vec{M}_i=\langle \vec{S}_{i} \rangle$ is the local magnetization. Magnetization and density are determined self-consistently for fixed total particle number $N$. In the following we assume zero temperature. The energy of the self-consistent solution is given by the sum over the lowest $N$  single-particle energies of the Hamiltonian~(\ref{eq:MF2}) plus the constant energy $E_0=U \sum_i( \vec{M}_i^2-\langle n_{i} \rangle^2/4)$.

An important subclass of self-consistent solutions 
are the ones with collinear magnetization where $M_y(i)=M_x(i)=0$ on all sites.
In particular,  the generic phases of the homogeneous Hubbard model\cite{Hirsch85}  have a collinear magnetization; either ferromagnetic $M_z(i)=M$, antiferromagnetic $M_z(i)=(-1)^i M$, or paramagnetic $M_z(i)=0$.
However, we will show that generally the combination of trapping potential and imbalance will lead to a non-collinear magnetization profile.

We are interested in the ground state for a given imbalance, characterized by the polarization $P=(N_\uparrow-N_\downarrow)/(N_\uparrow+N_\downarrow)$,  which is an experimentally controllable parameter\cite{zwierlein2006}. The imbalance is conserved, since the two components correspond to different internal states of the atoms (typically different hyperfine states) and transitions between these states are energetically forbidden  unless they are driven by additional lasers.
The single particle eigenstates of the Hamiltonian in Eq.~\eqref{eq:MF2} only have well-defined spin if the magnetization is collinear. Generally, an expectation $\langle S_z\rangle\neq 0$ can be tuned by spin-dependent chemical potentials or equivalently by a fictitious magnetic field in $z$-direction  $H_z= -B S_z$. 

The parabolic confinement will decrease the density away from the trap center. In a local density approximation, a cross section through the trap corresponds to a cut through the $(n,U)$ phase diagram at constant interaction $U$.  
Polarization can most easily be accommodated by ferromagnetism, but also antiferromagnetic and paramagnetic regions can account for finite imbalance. As discussed in the introduction, in a canted antiferromaget a spatially constant component aligned with the field is added to the alternating component perpendicular to the imbalance. The paramagnetic region can be partially polarized in the spirit of Pauli paramagnetism, where the polarization is proportional to the applied field. In the following we show that canted antiferromagnetic order is realized at half-filling, and we study how the imbalance is distributed between Mott plateau and edge as a function of interaction and imbalance. 
Self-consistent solutions of the Hubbard model on the two-dimensional square lattice~(\ref{eq:MF2}) have either a collinear or coplanar magnetization\cite{Verges91,gottwald2009}  and we can set $M_y=0$ without loss of generality. However, we note that enforcing vanishing global in-plane magnetization can lead to non-trivial three-dimensional topologies for the intrinsic ferromagnet\cite{Berdnikov09,LeBlanc09}, which will be discussed in Sec.\ref{Topology}.

\subsection{The homogeneous system at half-filling}\label{homogeneous}

\begin{figure}[ht!]
  \begin{center}
    \includegraphics*[angle=0,width=0.9\linewidth]{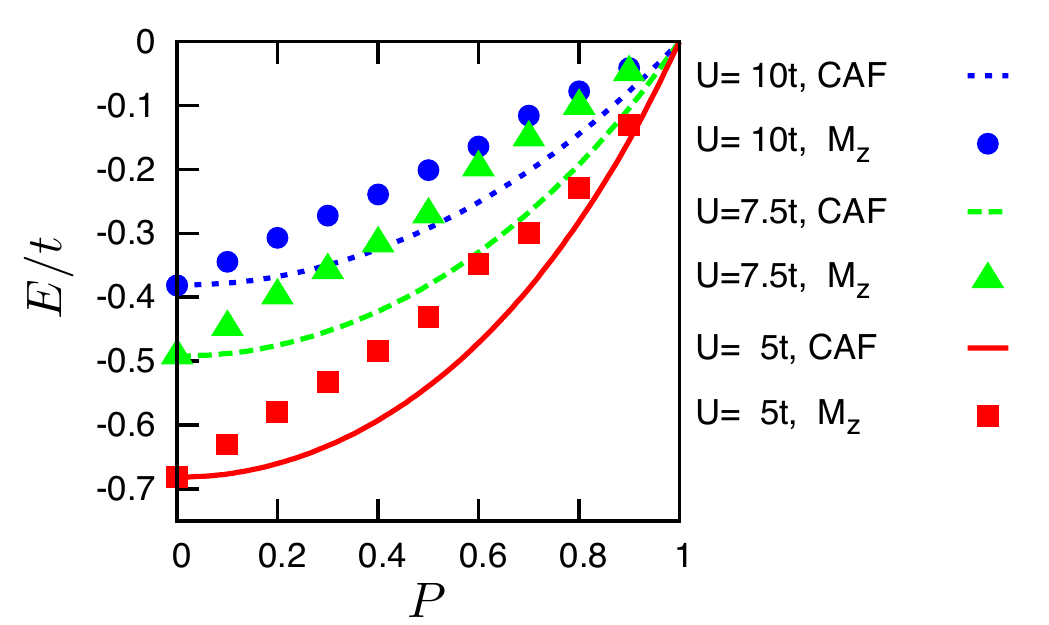} 
		    \caption{(Color online) Energy per particle as a function of polarization for the homogeneous Hubbard model at half-filling for various interaction strengths. Solutions with collinear magnetization ($M_z$) have higher energy than solutions with canted antiferromagnetic (CAF) order. Results obtained on a 20x20 lattice with periodic boundary conditions. For the canted antiferromagnet, the imbalance is fixed by a fictitious magnetic field in the $z$-direction, however, the Zeeman energy is not included in the plotted energies.}
\label{fig:Hom}
\end{center}
\end{figure}

Fig.~(\ref{fig:Hom}) shows the mean-field energies of canted and collinear solutions as a function of increasing imbalance for the homogeneous system at half-filling. A rough explanation of why the canted antiferromagnetic order is favored can be given within the mean-field Heisenberg model. Here the energy increases only quadratically with polarization for the canted order, but linearly with polarization for collinear magnetization. Since the solutions are the same at the extreme values  P=0 and P=1, the ground-state is always a canted antiferromagnet.

\subsection{Magnetization profile in the  trap}

\begin{figure}
  \begin{center}
   \includegraphics*[angle=0,width=0.9\linewidth]{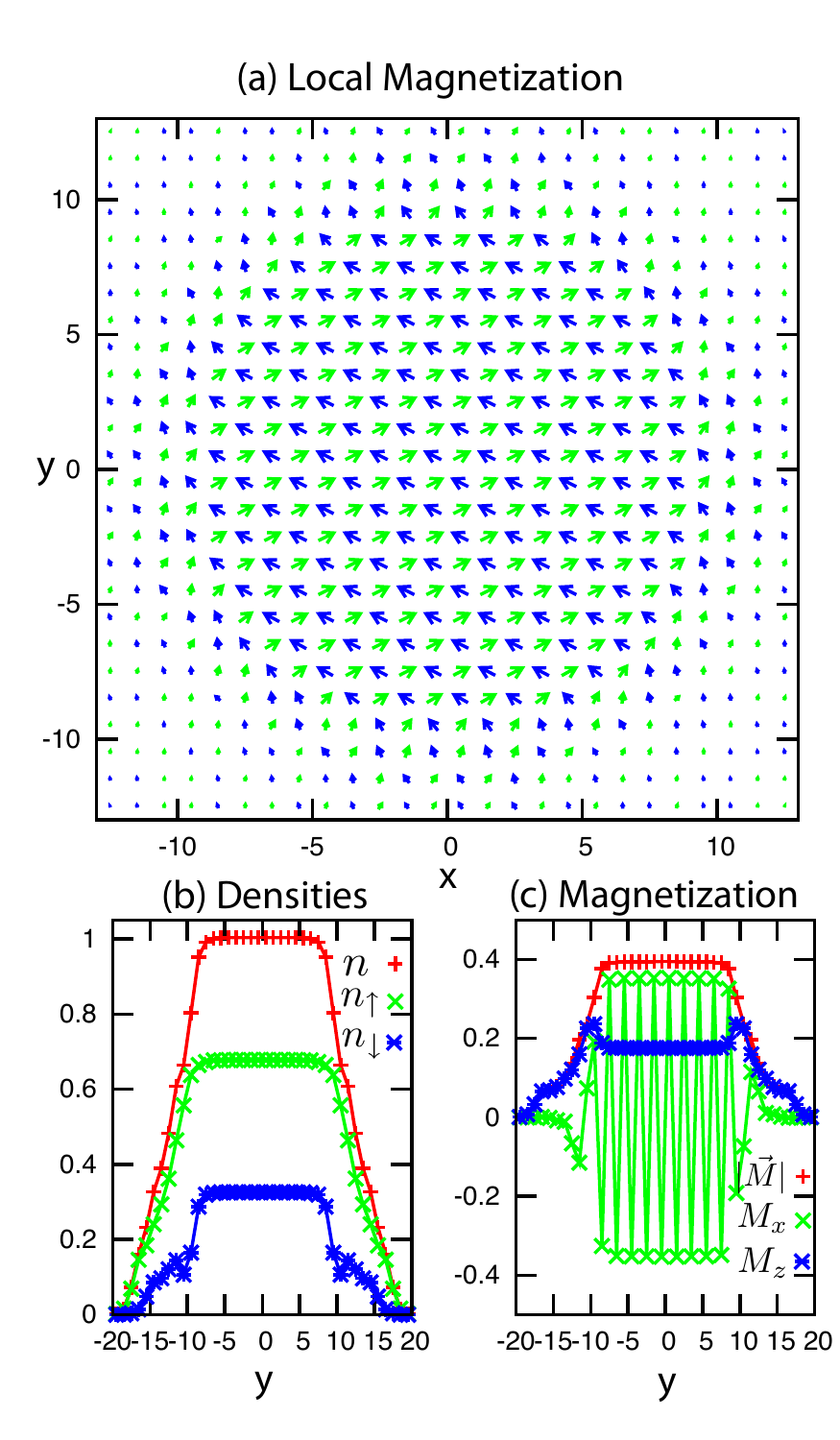} 
		   \caption{(Color online) Magnetic structure for $U=5t$, $P=0.5$, $N=540$, and $\alpha=0.02$ ($V_t=3.4t$).  (a) Local magnetization at the  lattice sites. Vertical (horizontal) component of arrows encodes  $M_z$ ($M_x$). Dark (blue) corresponds to negative values of $M_x$ and light (green) to positive values of $M_x$.   $x$ and $y$ denote spatial coordinates. (b) Spin-resolved densities along cross section at $x=0.5$.  c) Spin-components along cross section at x=0.5. The lattice size is 40x40.  }
\label{fig:Mag1}
\end{center}
\end{figure}

Fig.~(\ref{fig:Mag1}) shows an example of a typical magnetization profile of a self-consistent
solution at an intermediate interaction strength $U=5t$. For the chosen parameters, the interaction is strong enough to form a Mott plateau with $\langle n(i)\rangle = 1$ in the center. Furthermore, the trap strength, $\alpha=0.02t$, and the particle number, $N=540$, correspond to $V_t=3.4t$ which is smaller than the on-site interaction so that double occupancies are absent. 

\begin{figure}
\begin{center}
   \includegraphics*[angle=0,width=\linewidth]{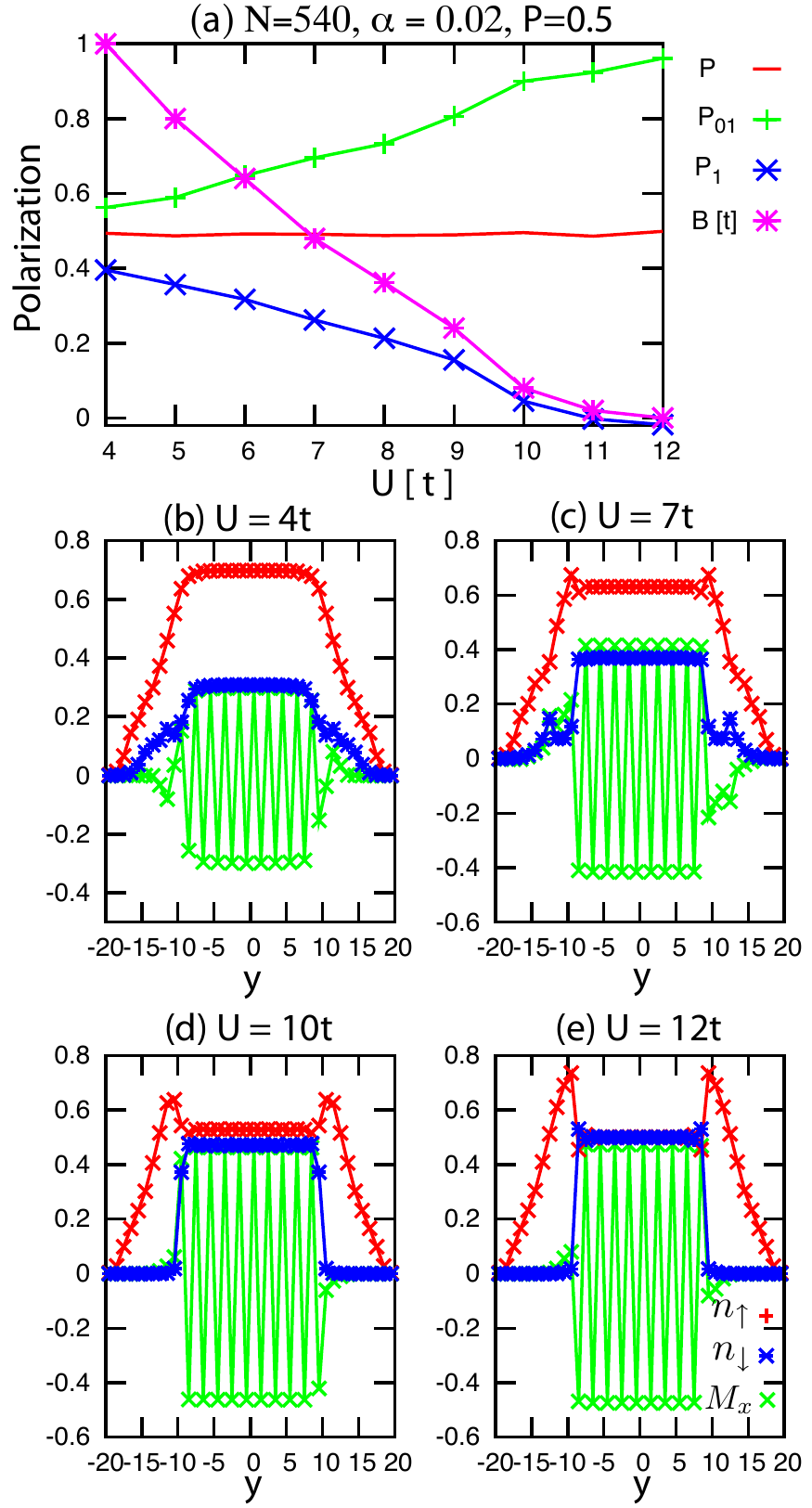} 
   \caption{(Color online) (a) Spatial distribution of imbalance as a function of interaction strength for constant global imbalance $P=0.5$. Increasing interaction increases polarization at the edge ($P_{01}$) and decreases polarization in the center ($P_1$). The fictitious magnetic field applied to fix the imbalance is shown with stars. (b)-(e) Cross-sections at $x=0.5$ of spin-resolved densities and $M_x$. Labeling is shown in (e).}
\label{fig:Int}
\end{center}
\end{figure}

\begin{figure}
  \begin{center}
    \includegraphics*[angle=0,width=\linewidth]{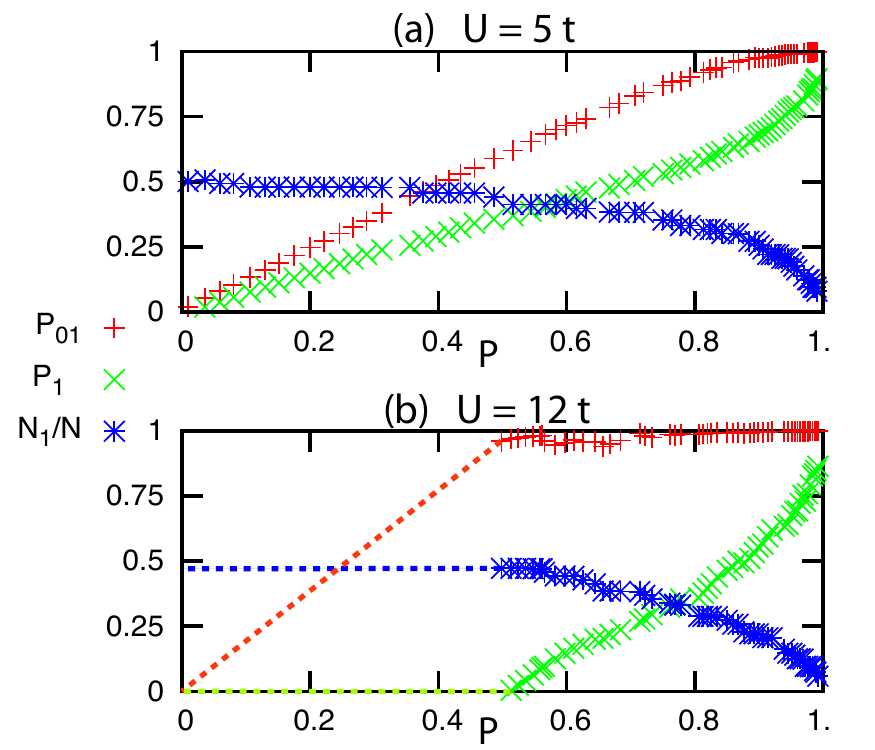} 
    \caption{(Color online) Spatial dependence of imbalance as a function of global imbalance.  $P_{01}$ denotes polarization at the edge. $P_{1}$ ($N_1$) denotes polarization (number of atoms) in the Mott plateau for (a) $U=5t$ and (b) $U=10t$. The dashed lines are explained in the text. All other parameters are as in Fig.~(\ref{fig:Mag1}).}
\label{fig:ImBal}
\end{center}
\end{figure}

\begin{figure}
  \begin{center}
   \includegraphics*[angle=0,width=\linewidth]{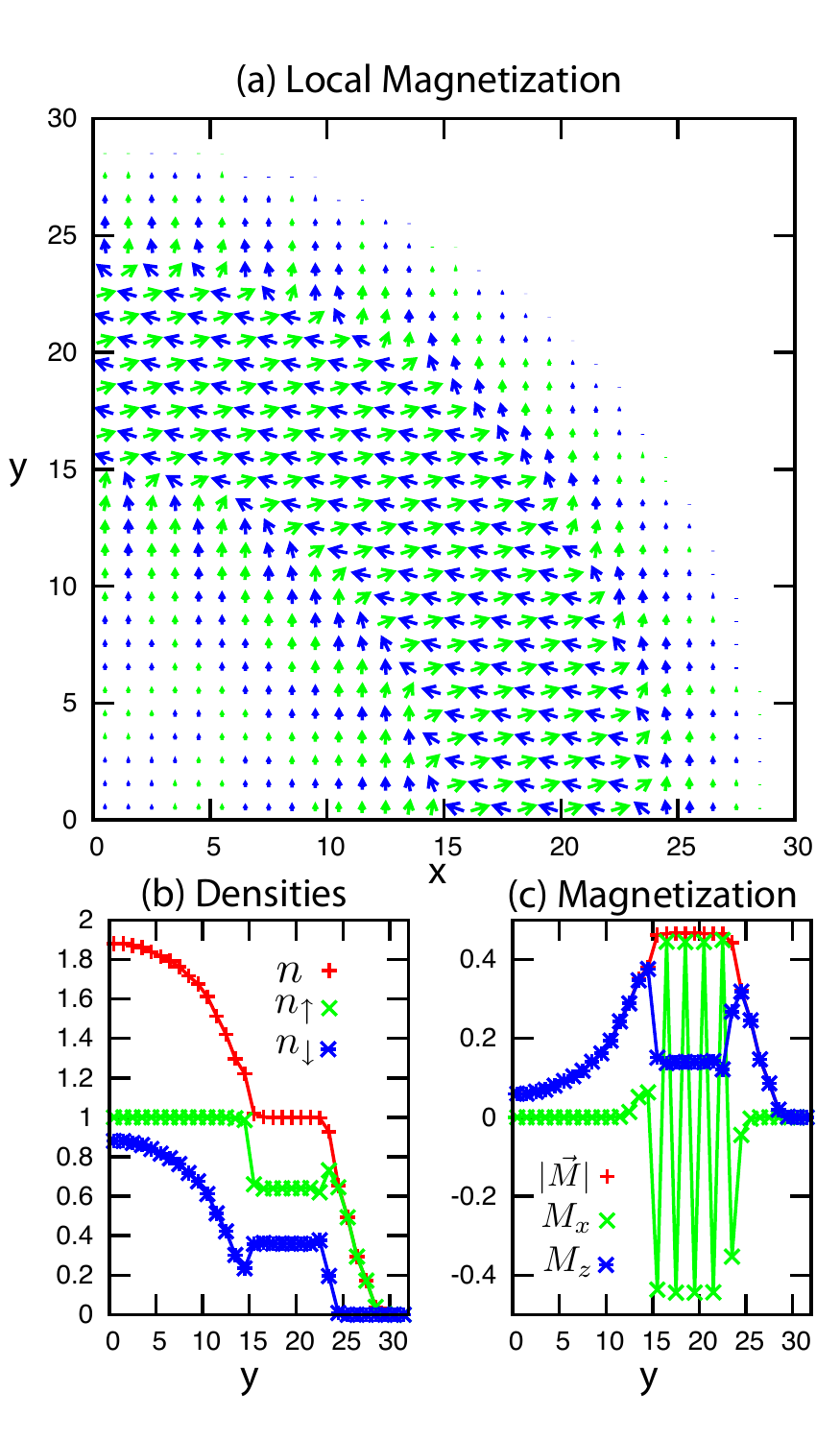} 
    \caption{(Color online) Same as Figure~\ref{fig:Mag1} but for $U=10t$, $N=2472$, $\alpha=0.02$, and $P=0.37$.  The lattice size is 64x64.}
\label{fig:Wedd}
\end{center}
\end{figure}

Within the Mott plateau we find canted antiferromagnetic order, 
as expected from the analysis of the homogeneous system.
The cross sections of the spin resolved densities and the local magnetization in panels (b) and (c) of Fig.~(\ref{fig:Mag1}) show that the edge is partially polarized and does not have antiferromagnetic order, although the $x$-component of the magnetization extends into the edge.

We now consider the distribution of a fixed imbalance for various on-site repulsions. Fig.~(\ref{fig:Int}) illustrates that increasing interaction moves the imbalance to the edge.
(we define the Mott plateau through $|n_i-1|<0.05$).
 Above a critical interaction strength (of order $U_c\approx10t$) the edge is fully polarized and the Mott plateau is a pure antiferromagnet. The maximum in the majority density at the border of the Mott plateau can be understood by recalling that in the homogeneous system for strong interactions, there is a 1st order phase transition between an antiferromagnet close to half-filling and a ferromagnet at finite doping\cite{Hirsch85}. By decreasing interactions below $U_c$, the canting in the Mott plateau increases and the polarization at the edge decreases.

Next we describe the magnetic structure as a function of the global polarization, $P$, keeping the other parameters fixed. For 
$U=5t$, the upper panel of Fig.~(\ref{fig:ImBal}) shows that both the polarization in the center with canted antiferromagnetic order and in the partially polarized edge increases linearly with the global polarization. 
The polarization at the edge is always larger than in the center until the Mott plateau disappears  close to full polarization.

We now discuss the case of strong interaction, i.e. $U>U_c$. Here the edge is intrinsically ferromagnetic. As shown in Fig.~(\ref{fig:Int}), at $U=12t$ the edge is already fully ferromagnetic in absence of any fictitious magnetic field that is otherwise used to fix a certain global imbalance. Given the total number of atoms in the trap, $N$, and the number of atoms in the edge, $N_{01}$, this defines a critical polarization $P_c=N_{01}/N$, which is $P_c\approx 0.5$ in Fig.~(\ref{fig:Int}) and Fig.~(\ref{fig:ImBal}).
Our mean-field approach predicts for $P<P_c$ and $U>U_c$ a spatially uniform ferromagnetic edge with a direction other than the $z$-direction. This implies a finite global in-plane magnetization.  However, as we discuss in detail in the next section such a solution which is forbidden by spin conservation, and the preferred ferromagnetic order in the edge will have nontrivial spin textures for $P<P_c$ and $U>U_c$.
For now we restrict our discussion to $P>P_c$ and $U>U_c$.  Then the ferromagnetic order at the edge points in the $z$-direction and the antiferromagnet in the Mott plateau is canted as shown in the lower panel of Fig~\ref{fig:ImBal}. 

We now increase the number of particles so that the center of the trap is more than half-filled. 
In agreement with  the symmetry of the homogeneous Hubbard model around half-filling, we find that the edge between the Mott plateau and double occupied sites shows similar features as the outer edge discussed above. 
Fig.~(\ref{fig:Wedd}) shows the magnetization profile and the spin-resolved densities. Here $V_t= 15.7$ which is larger than the chosen on-site interaction.  The Mott plateau is formed on a ring and has canted antiferromagnetic order. Moving away from the Mott ring, the antiferromagnetic order rapidly vanishes and the edge is strongly polarized. In fact, for this rather large value of $U$ we see a small maximum of the majority component at the outer edge and a minimum in the minority component at the inner edge.

\section{Non-trivial spin textures for $U>U_c$ }\label{Topology}

The Hartree-Fock calculation predicts that above a critical interaction strength $U_c$ the edge of the atom cloud turns ferromagnetic, even in absence of any imbalance or fictitious magnetic field.
In the previous section we defined a critical polarization, $P_c$, corresponding to a fully polarized ferromagnetic edge along the $z$-direction and an antiferromagnetic Mott plateau.
In this section we discuss qualitatively the magnetic structure for $U>U_c$ and $P<P_c$.

A cold atom experiment is prepared from a paramagnetic state with no optical lattice. 
Controlling the imbalance between the two fermion species, the initial state is characterized by
\begin{eqnarray}
\langle M_z \rangle&=& P N/2\,;\; \langle M_x \rangle=0=\langle M_y \rangle,\,\label{eq:bound}
\end{eqnarray} 
where $P$ is the polarization and $N$ the number of atoms. 
 Since there is no coupling between the effective spin degree of freedom and the rest of the experimental system, the same constraints apply in the presence of an optical lattice and with strong onsite interaction $U$.\cite{Berdnikov09,LeBlanc09}
This additional constraint is always fulfilled in our mean-field treatment except for $U>U_c$ and $P<P_c$, where a spatially uniform ferromagnetic edge is predicted with a direction other than the $z$-direction. However, such a solution leads to a finite global in-plane magnetization, which is forbidden by the boundary condition.
In order to fulfill Eq.~\eqref{eq:bound}, itinerant ferromagnetism in cold atom systems can have non-trivial topology as shown recently for balanced systems with filling factor less than unity everywhere~\cite{Berdnikov09,LeBlanc09}. 

In the following we discuss  the magnetic structure for $U>U_c$ and $P<P_c$, both for two-dimensional (2D) and three-dimensional (3D) systems. 
We look for magnetic structures that fulfill spin conservation~\eqref{eq:bound} and which show the  two prominent features predicted by the mean-field calculation, namely a) magnetic instabilities towards ferromagnetism in the compressible edge and antiferromagnetism in the Mott plateau and b) at the interface between Mott plateau and compressible edge, the orientation of the antiferromagnet and the ferromagnet should be perpendicular to each other.
Our qualitative analysis is based on the Ginzburg-Landau-type free energy functional (see  e.g. Ref.~\onlinecite{Berdnikov09} and~\onlinecite{LeBlanc09})
\begin{eqnarray}
E=\int \, d^2 r \, \frac{\rho}{2} |\nabla \vec{M}|^2 + \frac{\beta}{4}(|\vec{M}|^2 - |\vec{M}_0|^2)^2\,,\label{eq:Ginzburg}
\end{eqnarray}
where  $\rho$ is the positive stiffness constant, $M_0$ is the magnitude of the favored magnetization, and $\beta>0$ determines the cost of amplitude fluctuations. 
The favored spin texture for strong interactions, $U>U_c$, is determined by minimizing the total energy under the constraint of Eq.~(\ref{eq:bound}). 
In our qualitative analysis, we neglect that at the edge the system parameters in Eq.~\eqref{eq:Ginzburg} depend on the radius. This allows us to write the total energy of a spin structure, as a sum of three contributions: the energies of the spin structures at the edge, inside the Mott plateau and at the interface of both regions.  
We note that the energy scale related with the spin structure of the ferromagnetic edge is of the order $t$ and thus much bigger than the small superexchange $t^2/U$ that determines the spin structure in the Mott plateau. Therefore we first minimize the free energy of the intrinsically ferromagnetic edge. 
The remaining two energy terms describe the interface between ferromagnetic and antiferromagnetic order at the edge of the Mott plateau and the energy of the spin structure in the Mott plateau.
Based on the different scaling with the system size we argue that the interface term dominates for large systems.  While the interface term scales with $r_M^{D-1}$, where $r_M$ is the radius of the Mott plateau and $D$ denotes the dimension, the antiferromagnet scales like $\ln r_M$ in 2D and like $r_M \ln r_M$ in 3D, as we show below.
We minimize the interface term by choosing the orientation of the ferromagnetic and the antiferromagnetic order to be perpendicular to each other at the interface between Mott plateau and compressible edge.
In the following we discuss solutions, where the Mott plateau has no net imbalance. In fact, in the limit of large interactions $U\to\infty$, the superexchange $t^2/U$ vanishes, so that one could allow for a strong polarization of the edge, by polarizing the Mott plateau in the opposite way. As estimated in Appendix~\ref{app2} such a solution is however higher in energy for realistic interaction strengths.

\subsection{2D lattice}

\begin{figure}[h!]
   \includegraphics*[angle=0,width=0.9\linewidth]{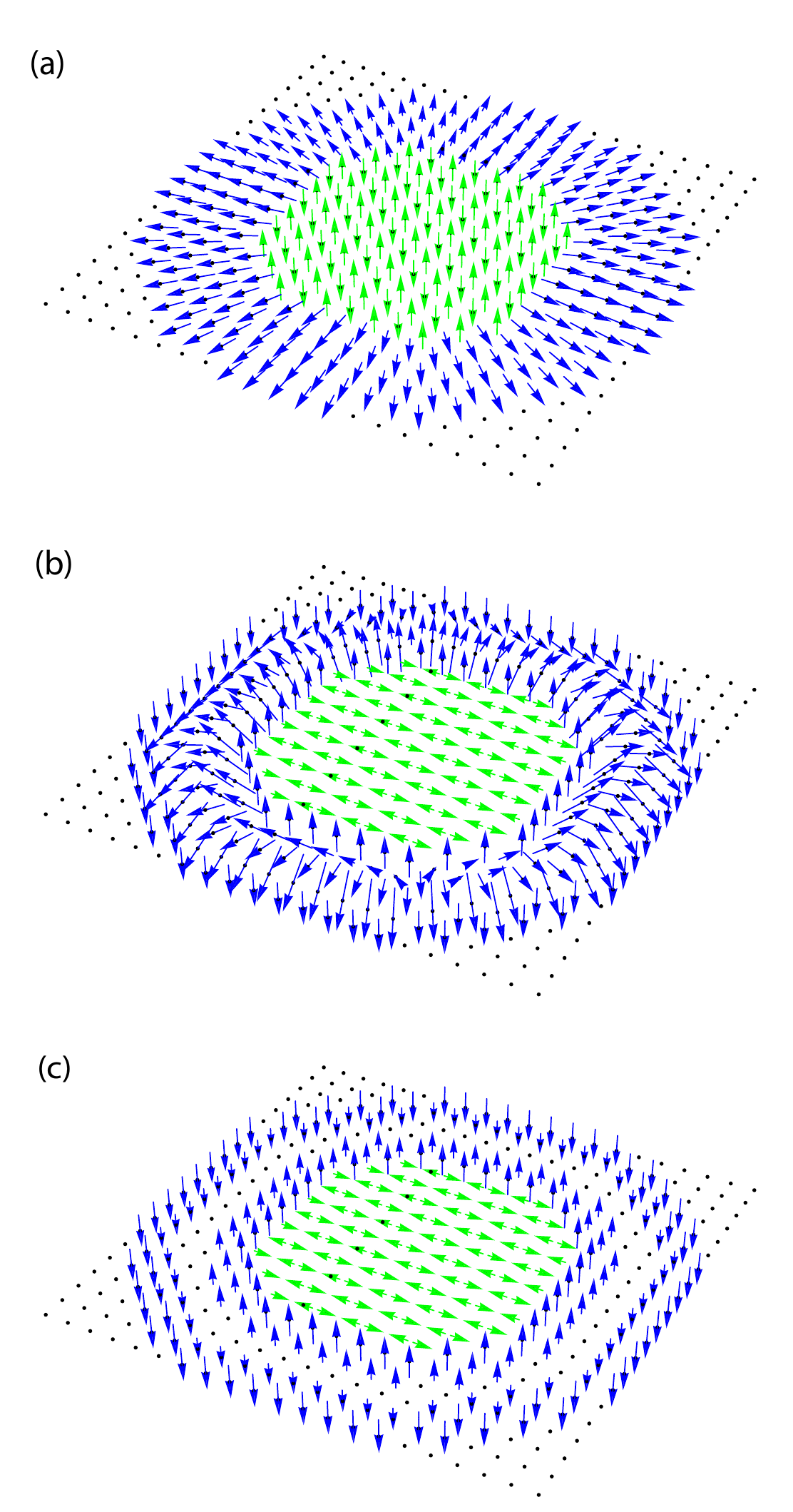}\\
     \caption{(Color Online) Illustration of possible magnetic structures at $U>U_c$ for balanced 2D systems with vanishing global magnetization. Dark 
     (blue) arrows indicate spin texture of the ferromagnetic edge. Light (green) arrows illustrate the magnetization in the antiferromagnetic Mott plateau. 
    Black dots indicate the regular 2D lattice.
    As discussed in the text the energetically favored solution is depicted in (a) and consists of a vortex structure at the ferromagnetic edge (here in the $xy$-plane) that is perpendicular to the antiferromagnetic ordering in the Mott plateau (here in the $z$-direction). (b) and (c)  illustrate a Skyrmion and domain wall structure in the ferromagnetic edge. }
\label{fig:Topology2Da}
\end{figure}

\begin{figure}[h!]
  \begin{center}
   \includegraphics*[angle=0,width=0.9\linewidth]{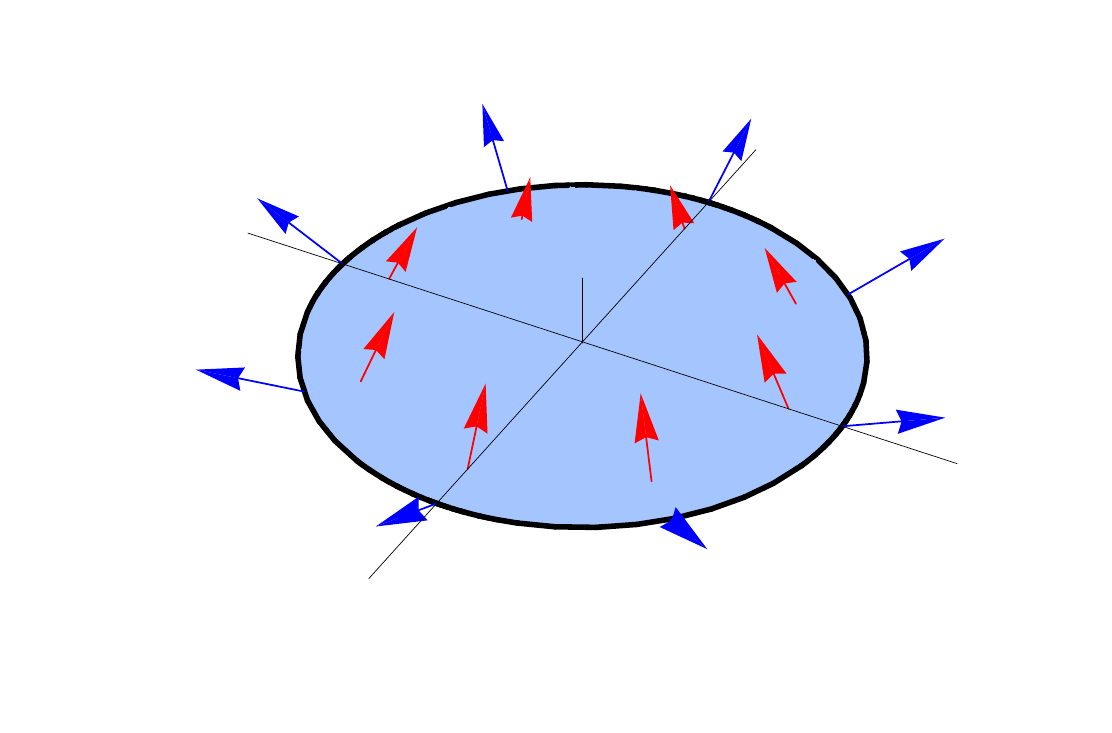}
      \caption{(Color Online) Illustration of magnetic structures in 2D for an imbalanced system with $U>U_c$ and $0<P<P_c$. Outer (blue) arrows indicate spin texture of the ferromagnetic edge which consists of a vortex in the $xy$-plane tilted towards the $z$-axis. Inner (red) arrows illustrate the orientation of the staggered magnetization in the antiferromagnetic Mott plateau which is perpendicular to the ferromagnetic order.       }
\label{fig:Topology2Db}
\end{center}
\end{figure}

We argue that (i) in presence of the Mott plateau a vortex structure for the ferromagnetic edge should be energetically favored as depicted in Fig~\ref{fig:Topology2Da} (a), and (ii) a finite imbalance should result in a vortex structure of the ferromagnetic order parameter in the $xy-$plane together with a small $z$-component, see Fig.~(\ref{fig:Topology2Db}). An important experimental consequence is a strong $z$-component of the antiferromagnetic order in the center, which is aligned perpendicular to the ferromagnetic order in the edge.
In Appendix~\ref{app} we derive the energy of the different topological orders of the ferromagnetic edge; vortex, domain wall, or Skyrmion. These structures are illustrated in Fig.~(\ref{fig:Topology2Da}).
It turns out that for realistic parameters, the vortex is lowest in energy. For finite imbalance the edge will then be described by a ferromagnetic vortex in the $xy-$plane and a constant $z$-component. The energetically preferred direction of the antiferromagnetic order in the Mott plateau is perpendicular to that of the ferromagnet at the edge. The antiferromagnet in the Mott plateau will therefore have a small in-plane magnetization forming a vortex, which grows with increasing imbalance, and a strong $z$-component as illustrated in Fig.~(\ref{fig:Topology2Db}).

\subsection{3D lattice}

\begin{figure}[h]
  \begin{center}
     \includegraphics*[angle=0,width=0.9\linewidth]{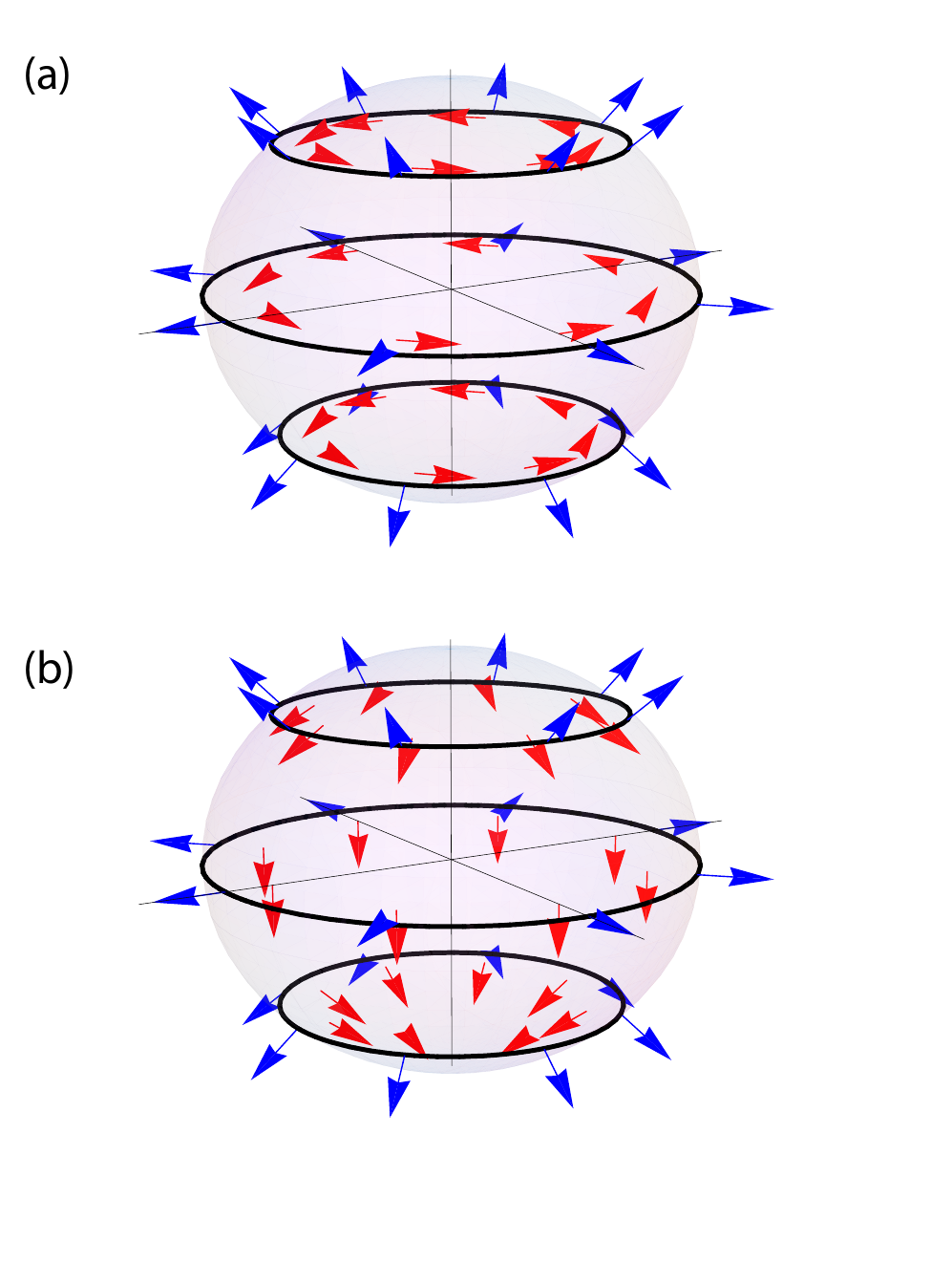}\\
     \caption{(Color Online) Illustration of possible magnetic structures in 3D for a balanced system with $U>U_c$. Outer (blue) arrows indicate the magnetization at the ferromagnetic edge and  inner (red) arrows illustrate the staggered magnetization in the antiferromagnetic Mott plateau. While the ferromagnetic edge always has a hedgehog structure the Mott plateau has either a planar vortex structure (a) or a 3D "spherical" vortex structure (b). While both structures have the same energy for the balanced system, the planar vortex is favored by finite imbalance, see Fig.\ref{fig:Topology3Db}  }
\label{fig:Topology3Da}
\end{center}
\end{figure}

\begin{figure}[h]
  \begin{center}
     \includegraphics*[angle=0,width=0.9\linewidth]{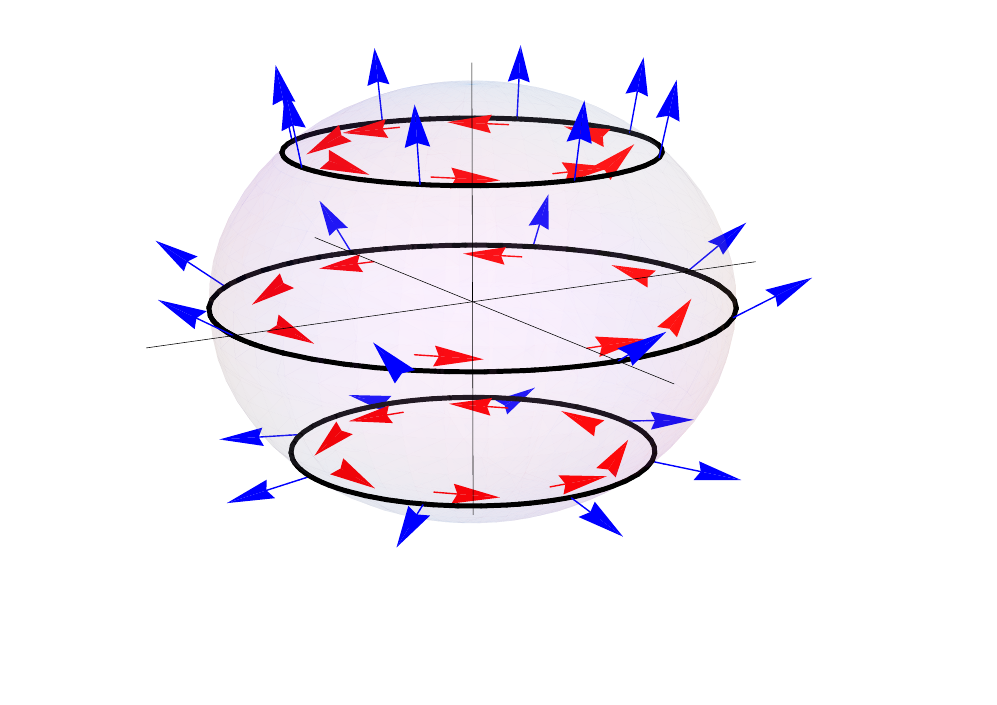}\\
    \caption{(Color Online) Illustration of magnetic structures in 3D for an imbalanced system with $U>U_c$ and $0<P<P_c$. Outer (blue) arrows indicate the magnetization at the ferromagnetic edge. Inner (red) arrows illustrate the staggered magnetization in the antiferromagnetic Mott plateau. Note that finite imbalance only deforms the hedgehog structure of the ferromagnet edge, while the antiferromagnetic order in the Mott plateau is unchanged, see Fig.~\ref{fig:Topology3Da} (a).}
\label{fig:Topology3Db}
\end{center}
\end{figure}

Similar arguments can be applied to a 3D system. Taking into account the boundary condition of vanishing global magnetization in balanced systems and by applying Eq.~\eqref{eq:Ginzburg}, one finds that the preferred structure of the ferromagnetic edge in a balanced system is a hedgehog\cite{Berdnikov09,LeBlanc09}. 
As shown in Appendix~\ref{app}, the energetically preferred antiferromagnetic order in the center should then be either a planar vortex structure with $\vec{M}_{AF}=\pm M_0 \vec{e}_\phi$ or a 3D spherical vortex $\vec{M}_{AF}=M_0 \vec{e}_\theta$, where $\vec{e}_\phi=(-\sin \phi, \cos \phi,0)$ and $\vec{e}_\theta=(\cos\theta\,\cos\phi, \cos\theta\,\sin\phi,-\sin\theta)$, are spherical unit vectors.

Both solutions are illustrated in Fig.~(\ref{fig:Topology3Da}). They guarantee that at the edge of the Mott plateau, where the antiferromagnetic order of the center of the trap has an interface with the ferromagnet order at the edge, the orientations of the antiferromagnet and the ferromagnet are perpendicular to each other. A violation of this requirement would cost an energy that scales with the area of the interface $r_M^2$.
Deformations of the perfect N\'eel order in the center of the trap, either in amplitude or phase, are minimized and the corresponding energy scales as $r_M\ln(r_M/a)$. For perfectly balanced systems, the vortex within the Mott plateau could lie in any plane.  Imbalance will deform the hedgehog leading to a net $z$-component, see Fig.~(\ref{fig:Topology3Db}). While this does not affect the energy of a vortex in the $xy-$plane it increases the energy for the vortices in other planes or for the spherical vortex.
Therefore we expect that for imbalanced systems in 3D with $U>U_c$, the antiferromagnetic order in the Mott plateau will form a planar vortex structure in the $xy-$plane as in Fig.~(\ref{fig:Topology3Db}).
In contrast to the 2D case where we expect a strong $z$-component of the antiferromagnetic order for $U>U_c$ and $P<P_c$, we expect a vanishing $z$-component in 3D.

\section{Discussion}\label{conclusion}
In this work we studied an interacting two-component Fermi gas  on a 2D and 3D cubic
  lattice subject to a parabolic external confinement. We analyzed the
  magnetic structure as a function of the repulsive
  interaction strength and spin imbalance.
  Applying an unrestricted Hartree-Fock calculation for a 2D system, we identified the critical interaction strength $U_c$ where the edge turns ferromagnetic and analyzed the spatial distribution of a finite imbalance between the two Fermi components for $U<U_c$. We found that the system has canted antiferromagnetic structure at half-filling with  antiferromagnetic ordering in the plane perpendicular to the imbalance and is partially polarized elsewhere. Fixing the global imbalance and increasing the interaction strength results in more imbalance flowing to the edge.  We expect the same qualitative behavior for 3D in that regime.
  
  In the second part of the work we gave a general discussion of the magnetic structure above $U_c$ both for 2D and 3D. We showed that spin conservation generally leads to nontrivial spin textures, both in the Mott plateau and at the edge. We predict that the edge has non-vanishing in-plane magnetization with a vortex structure in 2D and a hedgehog structure in 3D. We furthermore expect that for $U>U_c$ and small imbalance the antiferromagnetic order in the Mott plateau has a finite $z$-component in 2D, while in 3D a vanishing $z$-component of the antiferromagnetic order in the Mott plateau is predicted.
  
We expect our findings to have clear experimental signatures if temperatures below the N\'eel temperature can be reached.  A phase-contrast image \cite{zwierlein2006} showing the density of each component separately can test our predictions of a Mott plateau with ferromagnetic borders. Detection of a canted  antiferromagnet in the Mott plateau requires direct access to the order parameter. This can be achieved for instance through noise correlations \cite{altman2004} or by measuring the local magnetization\cite{vengalattore2007,Trotzky08,Greiner}. Additionally one can use Bragg spectroscopy\cite{Bragg1,Brag3} where the double unit cell of the antiferromagnet results in additional Bragg peaks. Furthermore the intensity of the additional Bragg peaks can then be used to measure the strength of the $z$-component of the antiferromagnet.

\section{Acknowledgements}
We thank David Pekker, Lode Pollet, Rajdeep Sansarma, David Jacob, Jan Zaanen, Ivar Zapata and Simon F\"olling  for illuminating discussion. Funding by the German Research Foundation under grants WU 609/1-1 (BW) and FR 2627/1-1 (LF), the NSF under grant DMR-0757145 (LF), and the Villum Kann Rasmussen foundation (NTZ) is gratefully acknowledged. The authors also acknowledge support  by MURI, DARPA-OLE program, CUA, and NSF Grant No. DMR-07-05472.

\appendix
\section{Ginzburg-Landau theory}\label{app}
Following Ref.~\onlinecite{Berdnikov09}, we apply a Ginzburg-Landau type description of the magnetism based on Eq.~\eqref{eq:Ginzburg} to analyze the magnetic structure for $U>U_c$, where the edge is intrinsically ferromagnetic.
By enforcing a vanishing global in-plane magnetization, the ferromagnetic edge acquires non-trivial topology. 
For the energy estimate we consider three energy contributions. The most relevant contribution is the free energy of the intrinsically ferromagnetic edge. Thereafter the contribution of the interface between ferromagnetic and antiferromagnetic order at the edge of the Mott plateau has to be taken into account, which is minimized by choosing the orientation of the ferromagnet and the antiferromagnetic to be perpendicular to each other. Finally the free energy of the antiferromagnetic Mott plateau has to be minimized.
 
We simplify our calculation by assuming constant parameters $\rho$, $\beta$, and $M_0$ in Eq.~\eqref{eq:Ginzburg}, thus neglecting a radial dependence of these parameters due to the trapping potential\cite{LeBlanc09}.We denote the radius of the atom cloud by $R_c$ and the radius of the Mott plateau by $r_M$.

\subsection{2D lattice} 

\begin{figure}
  \begin{center}
   \includegraphics*[angle=0,width=0.8\linewidth]{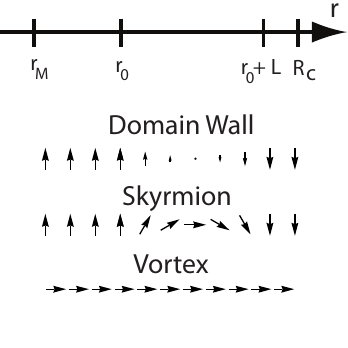} 
    \caption{Radial cross section through the ferromagnetic edge. Ferromagnetic edge starts at the border of the Mott plateau at $r=r_M$ and ends at $r=R_c$. While for the vortex the magnetization does not change in radial direction, both Skyrmion and Domain wall do change in radial direction within a ring defined by $r_0<r<r_0+L$.  Note that the radial component of the magnetization changes by $2\pi$ around the circumference while the $z$-direction is fixed.}
\label{fig:Topology2}
\end{center}
\end{figure}

In a 2D system we expect the magnetization at the edge to form a vortex-like structure.
Furthermore, we claim that for a small imbalance the vortex will lie in the $xy$-plane with a uniform magnetization component pointing in the $z$-direction. The energetically preferred direction of the antiferromagnetic order in the Mott plateau is perpendicular to that of the ferromagnet at the edge. At the interface, the antiferromagnet in the Mott plateau will have an in-plane magnetization forming a vortex and a $z$-component.
The lowest energy corresponds to the maximally allowed $z$-component of the antiferromagnetic order parameter thus minimizing the in-plane vortex. 

We now give quantitative arguments for the physics described above based on a comparison of the energies of a ferromagnetic edge with different topologies; either a vortex, a domain wall, or a Skyrmion as depicted in Figs~\ref{fig:Topology2Da} and ~\ref{fig:Topology2}.
First we discuss the balanced system. 
For the vortex the direction of magnetization is independent of radius but it rotates by $2\pi$ on each circumference. A particular realization of a vortex is $\vec{M}_{V}=M_0 \vec{e}_r$. However, for the balanced system there is global rotation invariance and the plane of the vortex is arbitrary. Using Eq.~\eqref{eq:Ginzburg}, the energy cost of a vortex is given by $E_{V}= \pi \rho M_0^2 \ln(R_c/r_M)$. Even in the absence of a Mott-plateau, the lattice spacing, $a_0$, gives a natural cutoff for the core energy leading to $E_{V}< \pi \rho M_0^2 \ln(R_c/a_0)$. A vortex naturally fulfills the requirement of vanishing global magnetization in all three spatial directions.

Another possibility is the formation of a domain wall. In the inner ring $r_M<r<r_0$ there is a uniform polarization (e.g. $\vec{M}=M_0 \vec{e}_z$) and within a finite region, $r_0<r<r_0+L$, the sign of the magnetization is inverted,
e.g. $\vec{M}=M_0(1-2(r-r_0)/L) \vec{e}_z$. In the outer ring, $r_0+L<r<R_c$, the magnetization points in opposite direction, e.g. $\vec{M}=-M_0 \vec{e}_z$. 
While the inner and outer rings have a perfect uniform ferromagnetic order, the domain wall is energetically costly due to the suppression of the amplitude of the order parameter. The energy cost is given by $E_{D}= \pi \rho M_0^2(r_0/L+1/2)(4+4L^2/(15\xi^2))$, with $\xi=\sqrt{\rho/(\beta M_0^2)}$ denoting the coherence length. $r_0$ and $L$ are not independent of each other but related by the condition of vanishing global magnetization. In absence of any Mott-plateau, $r_M=0$, the smallest allowed value is $r_0/L\approx 0.6$ which increases with $r_M$. Neglecting the term containing the coherence length we therefore obtain a lower bound for the energy of the domain wall: $E_{D}>\pi \rho M_0^2 4$. 

Finally, we estimate the energy of a Skyrmion. The magnetization is uniform (e.g. $\vec{M}=M_0 \vec{e}_z$)  in the inner ring, $r_M<r<r_0$, and then it rotates by an angle $a\pi$ around a local axis in a ring of width $L$, $r_0<r<r_0+L$, e.g. $\vec{M}=M_0\cos(\frac{r-r_0}{L}a \pi) \vec{e}_z+M_0 \sin(\frac{r-r_0}{L}a \pi)  \vec{e}_r$. 
For $a=1$, the magnetization in the outer ring is inverted, while for other angles it has a vortex structure, e.g. $\vec{M}=M_0\cos(a \pi) \vec{e}_z+M_0\sin(a \pi)  \vec{e}_r$. 
The Skyrmion interpolates between the inner and outer rings by tilting the order parameter, keeping the amplitude of the magnetization fixed in constrast to the domain wall where the amplitude is suppressed.  
In the region, $r_0<r<r_0+L$, the magnetization of the Skyrmion changes in the radial direction and along the circumference.
The radial dependence of the magnetization gives rise to an energy contribution given by $E_{S}= \pi \rho M_0^2(r_0/L+1/2)(a\pi^2)$. Again the variables $r_0$, $L$, and $a$ are not independent of each other but related by the condition of vanishing global magnetization. By minimizing this energy only, and neglecting the energy cost of the change of magnetization along the circumference, we get a lower bound for the Skyrmion energy: $E_{S}>\pi \rho M_0^2 (r_M/(R_c-r_M)+1/2)\pi^2$. 

According to these estimates the lower bound of the energy for the domain wall is larger than the total energy of the vortex for $r_M>\exp(-4) R_c\approx R_c/50$, and the lower bound for the Skyrmion is larger than the vortex energy for $r_M> \exp(-5) R_c\approx R_c/150$. In fact, the real minima for both Skyrmion and domain wall will be larger. 
Since the radius of the whole atomic cloud is about 50 lattice sites\cite{jordans2008,Mott2}, our conservative estimate 
shows that the vortex should be favored for practically any size of the Mott-plateau. We note that for the results shown in the main part of this paper $r_M/R_c\approx 1/2$. 
For the balanced system there is global rotation invariance and the plane of the vortex structure is arbitrary. However, in presence of a finite imbalance the energetically preferred magnetization will be a vortex in the $xy$-plane with a uniform ferromagnetic $z$-component. Assuming that the directions of the ferromagnetic edge and the antiferromagnet in the Mott-plateau are perpendicular, we expect the direction of antiferromagnet to have a large $z$-component.

\subsection{3D lattice} 
Minimizing the free energy in Eq.~\eqref{eq:Ginzburg}, one finds that the preferred structure of the ferromagnetic edge in a balanced 3D system is a hedgehog\cite{Berdnikov09,LeBlanc09}.  We now explain why we expect the antiferromagnetic order in the center to have a planar vortex structure  for $U>U_c$ and $P<P_c$.
First, at the edge of the Mott plateau, the preferred direction of the antiferromagnetic order is perpendicular to the orientation of the ferromagnet order in the edge. A violation of this requirement will cost an energy that scales with the area of the interface $r_M^2$.
In order to fulfill the boundary condition at the edge of the Mott plateau, the antiferromagnetic order in the trap center can neither have perfect N\'eel order nor a hedgehog configuration, since the latter needs to be oriented in the radial direction.
One possibility is to build up the 3D magnetization from the preferred 2D solution for each plane $z$, which is given by $\vec{M}_{AF}=M_0 \left\{z/r_M \vec{e}_\rho - [1-(z/r_M)^2]^{1/2} \vec{e}_z\right\}$, where $\vec{e}_z=(0,0,1)$ and $\vec{e}_\rho=(\cos(\phi),\sin(\phi),0)$ are cylindrical unit vectors.
However, this solution is not realized in 3D, since the change in the $z$-component of the magnetization between different planes costs a large energy that scales with the volume of the Mott plateau $E_{AF}\propto r_M^3/a^2$.
In fact, the preferred magnetic orders in the Mott plateau are either planar vortex structures like $\vec{M}_{AF}=M_0 \vec{e}_\phi$ or 3D solutions like $\vec{M}_{AF}=M_0 \vec{e}_\theta$, where $\vec{e}_\phi=(-\sin \phi, \cos \phi,0)$ and $\vec{e}_\theta=(\cos\theta\,\cos\phi, \cos\theta\,\sin\phi,-\sin\theta)$ are spherical unit vectors. For the balanced system these solutions have the same energy given by $E_{AF} \approx 4 \pi M_0^2 \rho r_M \ln(r_M/a)$.  However, imbalance will deform the hedgehog at the edge of the trap leading to a net $z$-component. Such a deformation increases the energy of these solutions except for a vortex in $xy-$plane.
Therefore we expect the magnetization profile in 3D for $U>U_c$ and small imbalance to be given by a (slightly deformed) hedgehog ferromagnet at the edge of the trap and an antiferromagnetic order with a vortex structure in the $xy-$plane in the center of the trap. In contrast with the 2D case, where we expect a strong $z$-component in the antiferromagnetic order for $U>U_c$ and $P<P_c$, we expect a vanishing $z$-component of the antiferromagnetic order in the Mott plateau in 3D.
 
\section{Polarizing the Mott plateau}\label{app2}

In section~\ref{Topology} we propose spin structures for $U>U_c$ and $P<P_c$ that minimize the total energy while fulfilling the  spin conservation~\eqref{eq:bound}.
The constraints~\eqref{eq:bound} prohibits the formation of a uniform ferromagnet at the edge of the trap if simultaneously the Mott plateau has an antiferromagnetic structure with zero net imbalance.
However, since the constraints~\eqref{eq:bound} apply to the whole system, one could imagine a system consisting of a fully polarized ferromagnetic edge and a Mott plateau strongly polarized in the opposite direction, such that the global imbalance is small or even zero.
We now justify why such solutions are energetically more costly than the ones proposed in  section~\ref{Topology}.

We therefore discuss the magnetic structure of a balanced Fermi gas in a 3D trap.
The system can be divided into a Mott plateau for  radius $r<r_M$ and an edge for radius $r_M<r<R_c$.
In section~\ref{Topology} we claimed that the ferromagnetic structure at the edge forms a hedgehog.
Applying the Ginzburg-Landau type free energy in Eq.~\eqref{eq:Ginzburg}, the energy of a hedgehog can be estimated. We strongly simplify our calculation by assuming a constant density at the edge.\cite{Berdnikov09} The
magnetitude $M_0$ of the ferromagnetic magnetization is therefore constant along with the stiffness, $\rho$, in Eq.\eqref{eq:Ginzburg}. We now estimate the energy of the ferromagnetic hedgehog as $E_F=8 \pi \rho M_0^2 (R_c-r_M)$. 
The stiffness 
of a homogeneous Fermi gas is given by $\rho=1/(12 k_F^2 \chi_0)=\hbar^2/(36 m\, n)$, where $k_F$ is the Fermi wavevector, $\chi_0$ the magnetic susceptibility, $m$ the mass of the fermions, and $n$ is the density (see e.g. Ref.~\onlinecite{LeBlanc09}). 
For sufficiently small densities the mass of a particle hopping between nearest neighbors in a 3D cubic lattice is given by $m=\hbar^2/(t a^2)$, where $a$ is the lattice constant and $t$ the hopping matrix element between nearest neighbor sites. The stiffness on a 3D cubic lattice is therefore given by $\rho\approx t a^2/(36 n)$ and the energy of the balanced hedgehog becomes $E\approx (2\pi/9) t (R_c-r_M)/(a^4 n)\simeq t N_E^{1/3}$, where $N_E$ is the number of atoms at the edge of the trap. 
This energy could be gained by uniformly polarizing the edge. 
However, due to the conservation of the total imbalance the Mott plateau would then also be polarized by $P_M=N_E/N_M$, where $N_M$ denotes the number of atoms in the Mott plateau. 
The corresponding cost in energy can be estimated as  $E_{AF}\approx P_M^2 2 N_M 4t^2/U=8t^2 N_E^2/(U N_M)$, where $2 N_M$ is the number of nearest neighbors in the Mott plateau and $4t^2/U$ is the superexchange. 
The energy cost of polarizing the Mott plateau is smaller than the energy gain of forming a uniform ferromagnetic edge if $U/t>36N_E^{5/3}/(\pi N_M)$. This is not satisfied for realistic particle numbers $N_E,N_M>10^3$ and $N_E/N_M\simeq 1$. 
We thus conclude that in 3D for $U>U_C$ and $P<P_c$ the Mott plateau is not significantly polarized. The magnetic structures that minimize the total energy and fulfill Eq.~\eqref{eq:bound} are therefore the ones presented in section~\ref{Topology}. We expect similar arguments to hold in 2D.

\end{document}